\documentstyle[12pt,epsfig]{article}

\def\tstrut{\vrule height2.5ex depth0pt width0pt} 
\def\er#1#2{\relax\ifmmode{}^{+#1}_{-#2}\else$^{+#1}_{-#2}$\fi}
\newcommand{\be}{\begin{equation}}
\newcommand{\bea}{\begin{eqnarray}}
\newcommand{\ee}{\end{equation}}
\newcommand{\eea}{\end{eqnarray}}

\begin{document}
%

\begin{center}
{\Large \bf Error estimates for $\pi\pi$
scattering threshold parameters in Chiral Perturbation Theory to two
loops.}\footnote{ Work supported by DGES PB98-1367 and by the Junta de
Andaluc\'{\i}a FQM0225.}

J. Nieves\footnote{e-mail: jmnieves@ugr.es} 
  and E. Ruiz Arriola\footnote{e-mail:earriola@ugr.es}
\\
[2em] Departamento de F\'{\i}sica Moderna, Universidad de Granada,
E-18071 Granada, Spain.\\

(7 April, 2000)
\end{center}

\begin{abstract}
Using the analysis of ChPT to two loops, we perform an error analysis
of the threshold and low energy parameters, based on the uncertainties
for the one loop low energy parameters and the resonance saturation
mechanism. Different sets of one loop low energy constants have been
considered. Thus, the predictive power of the effective field theory,
is quantified on the basis of the present experimental uncertainties.
\vskip.3cm 
\noindent
\centerline{\it 
}

\centerline{\it Keywords: Chiral Perturbation Theory,$\pi\pi$-Scattering, 
Error analysis. }
\end{abstract}

Chiral perturbation theory (ChPT) is the effective pion field theory
where the expansion parameter is $m^2 / (\sqrt{4\pi}f)^2 $, with $m$
the pion mass and $f$ the weak pion decay constant. At some finite
order an increasing number of undetermined parameters is generated
which can, so far, only be fixed by experimental data. Besides the
theoretical uncertainty introduced by the finite number of terms
considered in the expansion, there is an additional experimentally
induced uncertainty. The low energy parameters inherit such an
uncertainty, thus limiting in practice the predictive power of the
effective field theory; if the errors at a given order, say ${\cal O}
(m^{2n} / (\sqrt{4\pi}f)^{2n})$ are larger 
than the contributions of the next order
${\cal O} (m^{2n+2} / (\sqrt{4\pi}f)^{2n+2})$, the calculation of the latter
becomes useless, unless more accurate experiments are performed.

To make our points quantitative, we consider $\pi\pi$ scattering as
the prototype reaction to study the role played by chiral symmetry
breaking and the validity of the chiral expansion within QCD at low
energies. The scattering lengths and effective ranges for the isospin
$I=0,1,2$ have been computed in $SU(2)$ ChPT at tree level \cite{We66},
one loop \cite{GL84} and two loops \cite{mksf95,bc97}, in a power
series expansion
\begin{eqnarray}
a_{I J} &=&  a_{I J}^{\rm tree}+ a_{I J}^{\rm 1 \, loop}+a_{I J}^{\rm
2 \, loop}
+ \cdots  \\
b_{I J} &=&  b_{I J}^{\rm tree}+ b_{I J}^{\rm 1 \, loop}+b_{I J}^{\rm
2 \,loop} + \cdots  \, \, .
\end{eqnarray}
At tree level, the number of parameters involved are two: the pion
weak decay constant $f_\pi = 93.2$ MeV  and the pion mass $m=
139.6$ MeV.  For the purposes of the present discussion the
experimental error bars in these parameters can effectively be taken
to be zero. At the one loop level one has four, in principle
undetermined, parameters $ \bar l_1, \bar l_2, \bar l_3, \bar l_4 $.
At the two loop level the whole $SU(2)$ $\pi\pi $ amplitude
information can be gathered into six coefficients, which in
Ref.~\cite{mksf95,bc97} have been called $ b_1 , \dots, b_6 $.
Ideally, these parameters could be extracted from a direct low energy
analysis of $\pi\pi$ scattering experiments. The data are,
however, too poor in the low energy region and several methods have
been devised. Motivated by the success of the resonance saturation
hypothesis at scales $\mu \sim 0.5 - 1 $ GeV, at the one loop
level~\cite{egpr89}, the authors of Ref.~\cite{bc97} suggest to
estimate the bulk of the two loop corrections through resonance
saturation.  This approach requires the values for the one loop $\bar
l$'s deduced from some clean source.  In Ref.~\cite{bc97} two
different sets of ${\bar l}$ parameters have been considered: i) by
taking the $\bar l_1 $ and $\bar l_2$ parameters obtained from
$K_{l4}$ form factors using a dispersive one loop calculation for
three flavors \cite{bcg94} (Set {\bf I} in~\cite{bc97}), and ii)
fixing $\bar l_1 $ and $\bar l_2$ to the values deduced from $D-$wave
$\pi\pi$ scattering lengths, at order ${\cal O}(p^6)$ (Set {\bf II}
in~\cite{bc97}). Unfortunately, no error analysis has been undertaken
in Ref.~\cite{bc97}. On the other hand, the estimates based on
$K_{l4}$ analysis, can now be improved thanks to a recent study of
this process at two loop accuracy~\cite{abt99}. Other possibility is
to extract the one loop low energy constants, together with their
errors, from Roy sum rules, saturated by the high energy behavior of
$\pi\pi$ scattering~\cite{gir97}. In conjunction with resonance
saturation, this procedure also yields an error analysis of threshold
parameters~\cite{gir97}.

An error analysis from the point of view of the predictive power of
ChPT is missing, and it is the main subject of the present work.  We
assume that primary quantities, coming from a $\chi^2-$fit or direct
experimental measurements, are Gauss distributed (with or without
correlations), as we learn from elementary statistics. We
propagate errors by means of Monte Carlo simulations, keeping always
statistical correlations between all parameters entering in a given
derived quantity. We use $10^4$ samples, and to quote errors we always
use a 68\% confidence level around the central value. Since the
out-coming threshold parameters distributions are not gaussians, we
take in this way into account possible skewness in the distributions.
By using the Monte Carlo method we avoid summing errors in
quadratures, which would be incorrect for statistically correlated
quantities, and we do not have to use any complicated covariance
formula.

Let $a^{(n)}$ and $\Delta
a^{(n)}$ be the n-loop central value and error of an observable. Thus,
to be {\it predictive} and {\it convergent} at the $n-$loop level one
ought to have the relation,
\begin{equation}
\Delta a^{(n)} << a^{(n+1)} << a^{(n)} . 
\end{equation} 

In order to establish these necessary requirements from a quantitative
point of view we will adopt the currently accepted scheme of assuming
that resonance saturation yields the bulk of the two loop contribution
at scales $0.5-1.0$ GeV. This induces a scale ambiguity in the
two--loop part of the $b-$parameters, which we implement by providing the scale
with an error,
\begin{equation} 
\mu = 750 \pm 250 \, {\rm MeV} \, .
\label{eq:scale}
\end{equation}   
Moreover, since one does not expect resonance saturation to be
exact, we provide these parameters with $100\%$ uncertainty and take,
in the notation of Ref.~\cite{bc97}, the values
\begin{eqnarray}
& & 
r_1 = -0.6 \times 10^{-4} \times (1 \pm 1) \quad 
r_2 = 1.3  \times 10^{-4} \times (1 \pm 1) \quad  
r_3 = -1.7  \times 10^{-4} \times (1 \pm 1) \quad \nonumber 
 \\  
& & 
r_4 = -1.0  \times 10^{-4} \times (1 \pm 1) \quad 
r_5 = 1.1  \times 10^{-4}  \times (1 \pm 1) \quad  
r_6 = 0.3   \times 10^{-4} \times (1 \pm 1)  \label{eq:res}
\end{eqnarray}
at the scale $\mu$ given in Eq.~(\ref{eq:scale}). In addition to these
numbers, one needs estimates of the one loop low energy parameters
$\bar l_{1,2,3,4}$. Through all this work we fix
\begin{eqnarray} 
& &
  \bar l_3 = 2.9 \pm 2.4 \, , \quad
  \bar l_4 =  4.4 \pm 0.3 \, , 
\end{eqnarray} 
as determined from the study of $SU(3)$ breaking effects and the
scalar form factor in Refs.~\cite{GL84} and~\cite{bct98} respectively.
There is less consensus regarding the values of $\bar l_1 $ and $\bar
l_2 $, therefore we consider three possibilities. We give below the
central values, errors and linear correlation coefficients for $\bar
l_1 $ and $\bar l_2$ in each case, in order to characterize the full
statistical distributions, deduced from different primary
distributions, which will be used later to calculate threshold and low
energy parameters.

\begin{itemize}

\item \underline{Set {\bf I} $K_{l 4}$ form factors:}
Ref.~\cite{abt99} allows to determine $\bar l_1 $ and $\bar
l_2$ from a two loop analysis of $K_{l4}$ decays.  The analysis is
done in terms of the scale dependent $SU(3)$ low energy constants $L_i^r
$ at the $\rho$ mass scale, $ m_\rho = 770 $ MeV, and central
values for $\bar l_1 $ and $\bar l_2$ are given. The relation between
the $SU(2)$ and $SU(3)$ low energy constants is known at one
loop~\cite{gl85} and thus the determination of the $\bar l_1 $ and
$\bar l_2$ parameters in~\cite{abt99} suffers from two loop $SU(3)$
uncertainties. From the remarks of that work a strong anti--correlation
between $\bar l_1 $ and $\bar l_2$ can be inferred, although the
corresponding correlation matrix is not given. We will take here two
extreme view points, total de--correlation and total anti--correlation. As
we will see, the totally anti--correlated case seems to induce smaller
errors in the scattering lengths as compared to the totally
de--correlated case. We will call these two choices Set {\bf Ia} and
Set {\bf Ib} respectively. We will also examine here a partial
anti--correlation scenario (Set {\bf Ic}) which models more
accurately the limited statistical information  provided in
Ref.~\cite{abt99}. More details are given in the Appendix.

For the three sets, we take $L_1^r = 0.52 \pm
0.23$, $L_2^r = 0.72 \pm 0.24$ and $L_3^r = -2.69 \pm 0.99$  (main fit
in Table 1 of Ref.~\cite{abt99}). For Set {\bf Ia}, we assume that 
the three statistical distributions are de--correlated, whereas  for
Set {\bf Ib}  [Set {\bf Ic}] assume these are correlated as specified in
Eq.~(\ref{eq:corr})  [ Eq.~(\ref{eq:Cmatrix}) with $r_{13} =
r_{23}=-r_{12} = -0.85$]. By means of a Monte Carlo simulation we generate the 
statistical distributions for the $\bar l_1$ and $\bar l_2$
parameters, and thus we get
\begin{eqnarray} 
& &
  \bar l_1 = 0.3 \pm 2.1 \, , \quad
  \bar l_2 =  4.77 \pm 0.45 \, , \quad 
  r(\bar l_1,\bar l_2 ) = \phantom{-}0\, ,\phantom{.70}   \qquad {\rm Set} \, \, {\bf Ia } \\
& &
  \bar l_1 = 0.3 \pm 1.0 \, , \quad
  \bar l_2 =  4.77 \pm 0.45 \, , \quad
  r(\bar l_1,\bar l_2 ) = -1\, ,\phantom{.70}  \qquad {\rm Set} \, \, {\bf Ib } \\
& &
  \bar l_1 = 0.3 \pm 1.2 \, , \quad
  \bar l_2 =  4.77 \pm 0.45 \, , \quad 
   r(\bar l_1,\bar l_2 ) = -0.69 \, , \qquad {\rm Set} \, \, {\bf Ic } 
\end{eqnarray}
being $r$ the linear correlation coefficient. The errors quoted above
are clearly low bounds, because they do not account for any systematic
effects, in particular those induced by the ${\cal
O}(p^6)-$corrections to the relations between the two-- and
three--flavor low energy constants\footnote{Note that in the second
entry of Ref.~\protect\cite{abt99} there is an inconsistency. In
Eq.~(6.24) an error of $\pm 1.0$ for ${\bar l}_2$ is quoted. That
error is not compatible with having ${\bar l}_2 = 192 \pi^2 L_2^r -
(1+\ln( m_K^2/\mu^2) + 8 \ln (m^2/\mu^2) )/8 $, as deduced from
Eq.~(6.23) of that reference, and an error for $L_2^r$ of $\pm
0.24\times 10^{-3}$. The previous formula gives an error for ${\bar
l}_2$ of $ \pm 192 \pi^2 \times 0.24\times 10^{-3} = \pm 0.45$ and it
is obviously independent of the possible correlations between the
$SU(3)$ $L_1^r,L_2^r$ and $L_3^r$ parameters, because it only involves
$L_2^r$. This is in agreement with our results. The problem in
Ref.~\protect\cite{abt99} is that the error definition for the
$SU(2)-$parameters, ``projections on the relevant variable of the 68\%
confidence level domain'', is not consistent with that adopted for the
$SU(3)-$parameters. For these latter ones, the standard and
traditionally accepted $\chi^2-$errors are given. The method of the
projections would lead to significantly different errors for
$L_{1,2,3}^r$, as can be appreciated from the figures shown in
Ref.~\protect\cite{abt99}. In anycase, this projection method is not
standard. The {\it standard} and complete procedure consists of giving
the full correlation matrix deduced from the $\chi^2-$fit. We ignore
to what extent the errors on derived quantities given in
Ref.~\protect\cite{abt99} are affected by this
inconsistency. }. Estimates for the systematic errors are not given in
Ref.~\cite{abt99} either. Notice that correlations among $L_{1,2,3}^r$
do not affect to the error in ${\bar l}_2$ (see footnote 4).

\item \underline{ Set {\bf II} $D-$waves:}
The method of Ref.~\cite{bc97} allows to define another parameter set.  If the
$D-$wave $\pi\pi$ scattering lengths for isospin $I=0$ and $I=2$,
$a_{02}$ and $a_{22}$ respectively, are fixed at the two loop level
one gets parameter Set {\bf II},
\begin{equation}
  \bar l_1 = -0.8 \pm 4.8 \, , \quad
  \bar l_2 =  4.5 \pm 1.1 \, , \quad
  r(\bar l_1,\bar l_2 ) = -0.75 \, , \qquad {\rm Set} \, \, {\bf II } 
\end{equation}
To obtain $\bar l_1$ and $\bar l_2$ and their errors, we have
propagated the errors in $\mu$, $a_{02}$, $a_{22}$, $\bar l_3$, $\bar
l_4 $ and the resonance parameters of Eq.~(\ref{eq:res}) in the
formula for the $D-$wave scattering length given in
Ref.~\cite{bc97}. This procedure generates a correlation between $\bar
l_1 $ and $\bar l_2 $ which has to be taken into account when
calculating errors in quantities depending on the previous
parameters. Note that our central values for $\bar l_1$ and $\bar l_2
$ numbers are not exactly the ones quoted in Ref.~\cite{bc97} since
they take $\mu=1$ GeV.  Furthermore, in Ref.~\cite{bc97} no error
estimates are quoted for $\bar l_1$ and $\bar l_2 $. If we had taken
$\mu = 1.00 \pm 0.25 \,{\rm GeV} $, we would have obtained $ \bar l_1 =
-1.5 \pm 5.8 $ and $ \bar l_2 = 4.5 \pm 1.1 $, in agreement with their
quoted central value.

\item \underline{ Set {\bf III} Roy sum rules:}
Finally, using the method of Roy sum rules, Ref.~\cite{gir97},
another parameter set has been obtained. We
call it Set {\bf III},  
\begin{equation}
  \bar l_1 = -0.9 \pm 1.2 \, , \quad \bar l_2 = 4.34 \pm 0.25 \, , \quad
  r(\bar l_1,\bar l_2 ) = -0.22 \, , \qquad {\rm Set} \, \, {\bf III } 
\end{equation}
Both central values and errors quoted above do not agree with those
given in Eq.~(21) of Ref.~\cite{gir97}. The difference on the central
values is due to the fact that in this latter reference all resonance
parameters are set to zero at 1 GeV ($r_i(1 {~\rm GeV}) = 0 \pm
2\times 10^{-4}$). To obtain the errors we have propagated the errors
in the resonance saturation scale $\mu$, $\lambda_1$, $\lambda_2$
(parameters given in Eq.~(2) of Ref.~\cite{gir97}), $\bar l_3$, $\bar
l_4 $ and the resonance parameters of Eq.~(\ref{eq:res}) when
Eqs.(6-7) of Ref.~\cite{gir97} are inverted. This, again generates a
correlation between $\bar l_1 $ and $\bar l_2 $ as in the previous
cases\footnote{Errors differ from those quoted in Ref.~\cite{gir97},
because of the different treatment considered here (error on the
resonance saturation scale, different choice, both for central values
and errors, for the resonance parameters but also because there is a
numerical mistake in that work, which we correct here. The error
analysis of Eqs.~(16) and (17) of that work yields $\bar l_1 = -0.37
\pm 0.95 \pm 1.27 $ and $\bar l_2 = 4.17 \pm 0.19 \pm 0.33 $, which
errors do not agree with those quoted in Eq.~(20) of that reference. }. The $\lambda_1$,
$\lambda_2$ parameters suffer from sizeable systematic
uncertainties. Those have been estimated in the second entry of
Ref.~\cite{mksf95} and turned out to be of comparable magnitude, when
not bigger, than the statistical fluctuations, and as a consequence
statistical correlations get washed out. This justifies the use of
de--correlated distributions for these two parameters both in
Ref.~\cite{gir97} and in the present work.
\end{itemize}
\begin{figure}
\centerline{
\epsfig{figure=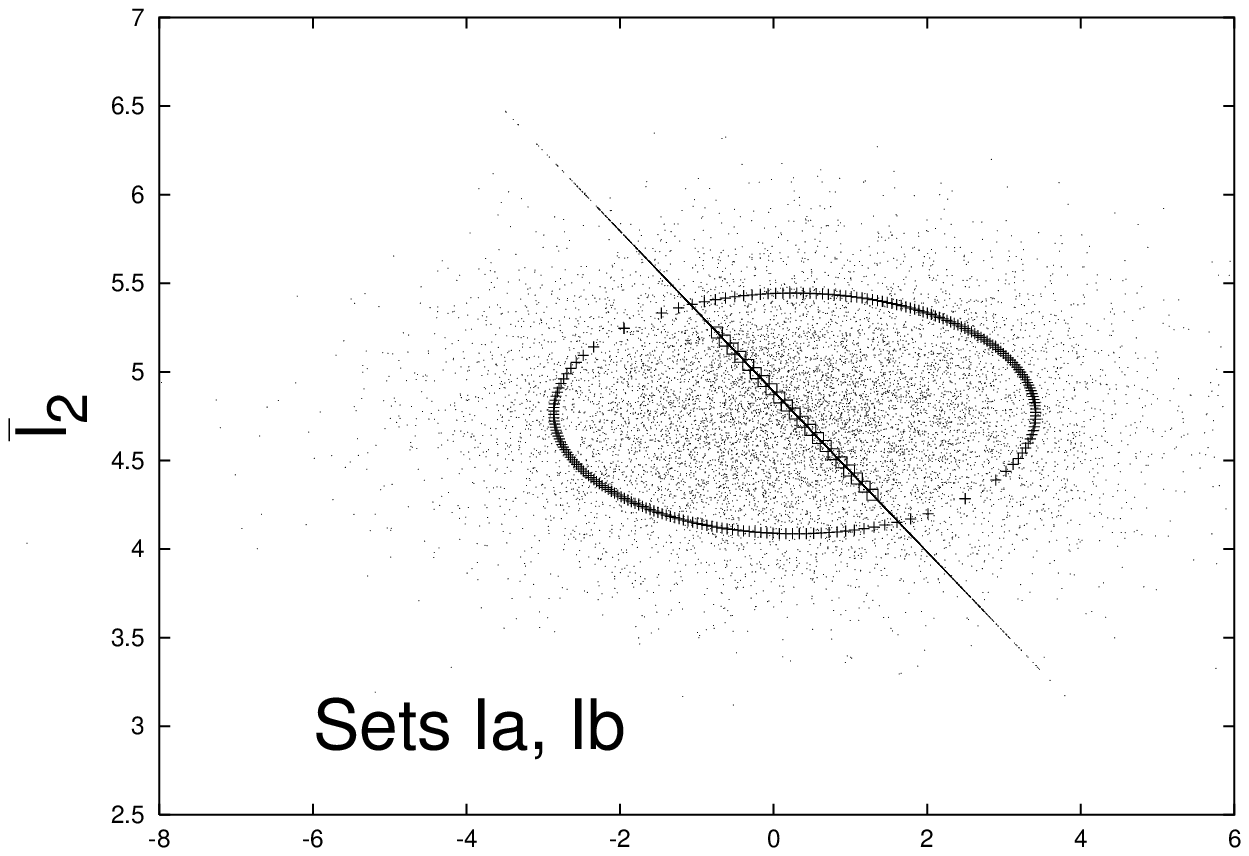,height=6cm,width=8.5cm}
\hspace{.5cm}
\epsfig{figure=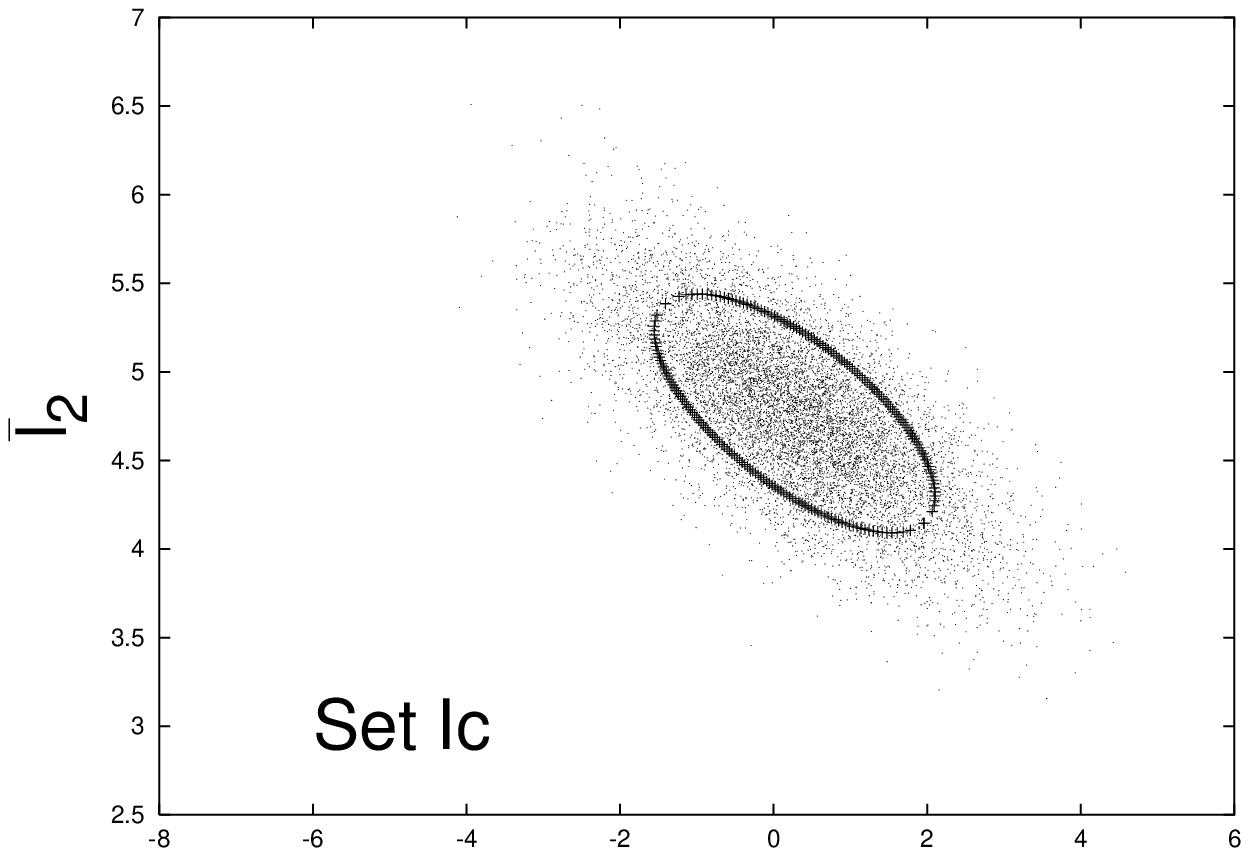,height=6cm,width=8.5cm}}
\centerline{
\epsfig{figure=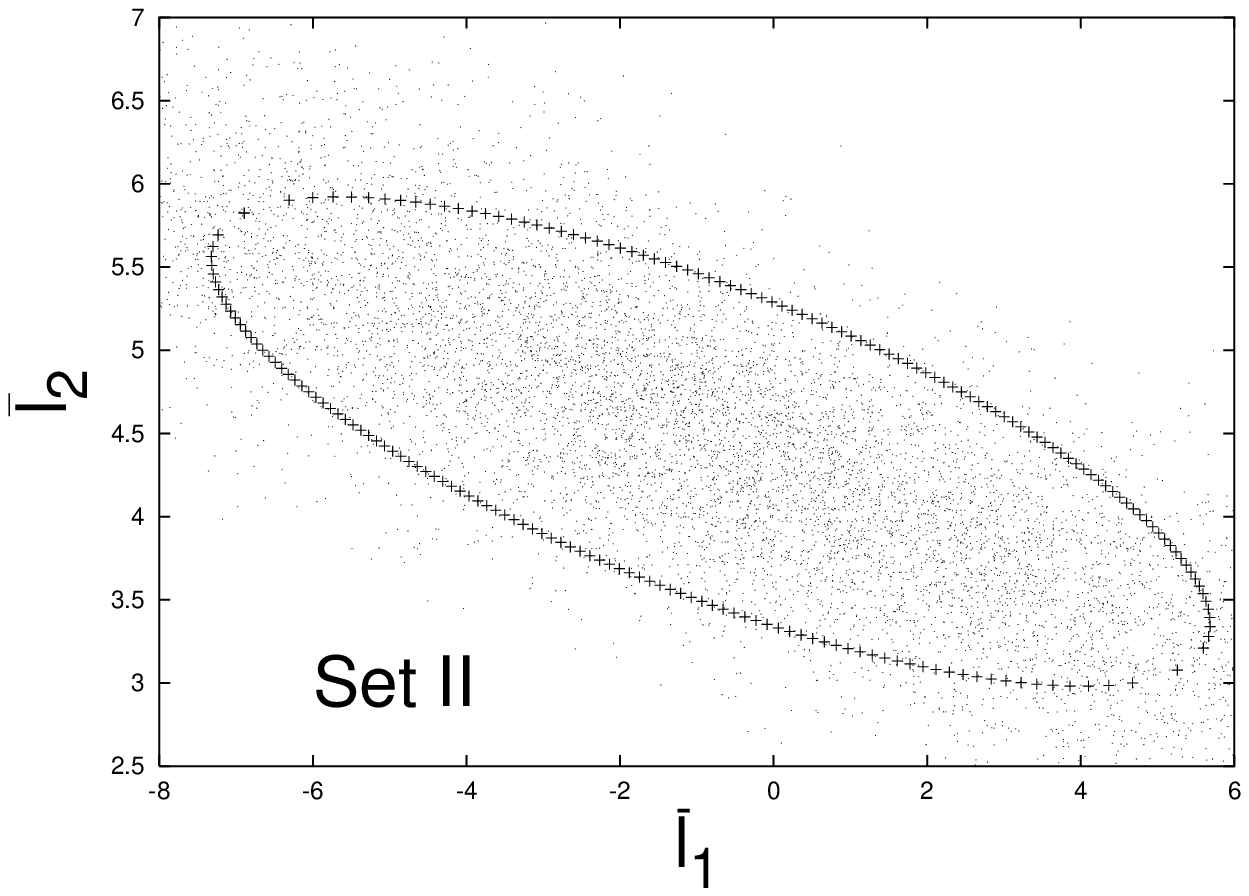,height=6cm,width=8.5cm}
\hspace{.5cm}
\epsfig{figure=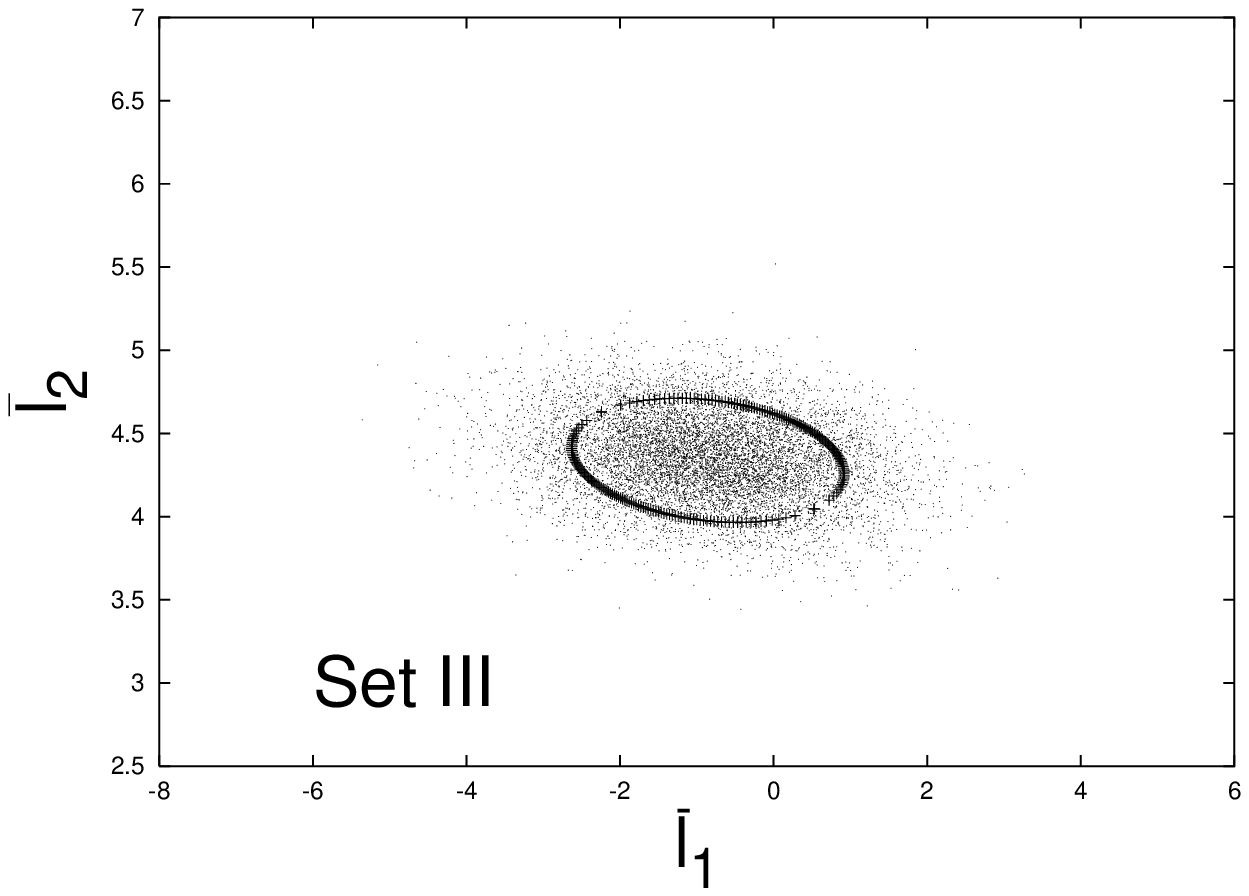,height=6cm,width=8.5cm} }
\centerline{
\epsfig{figure=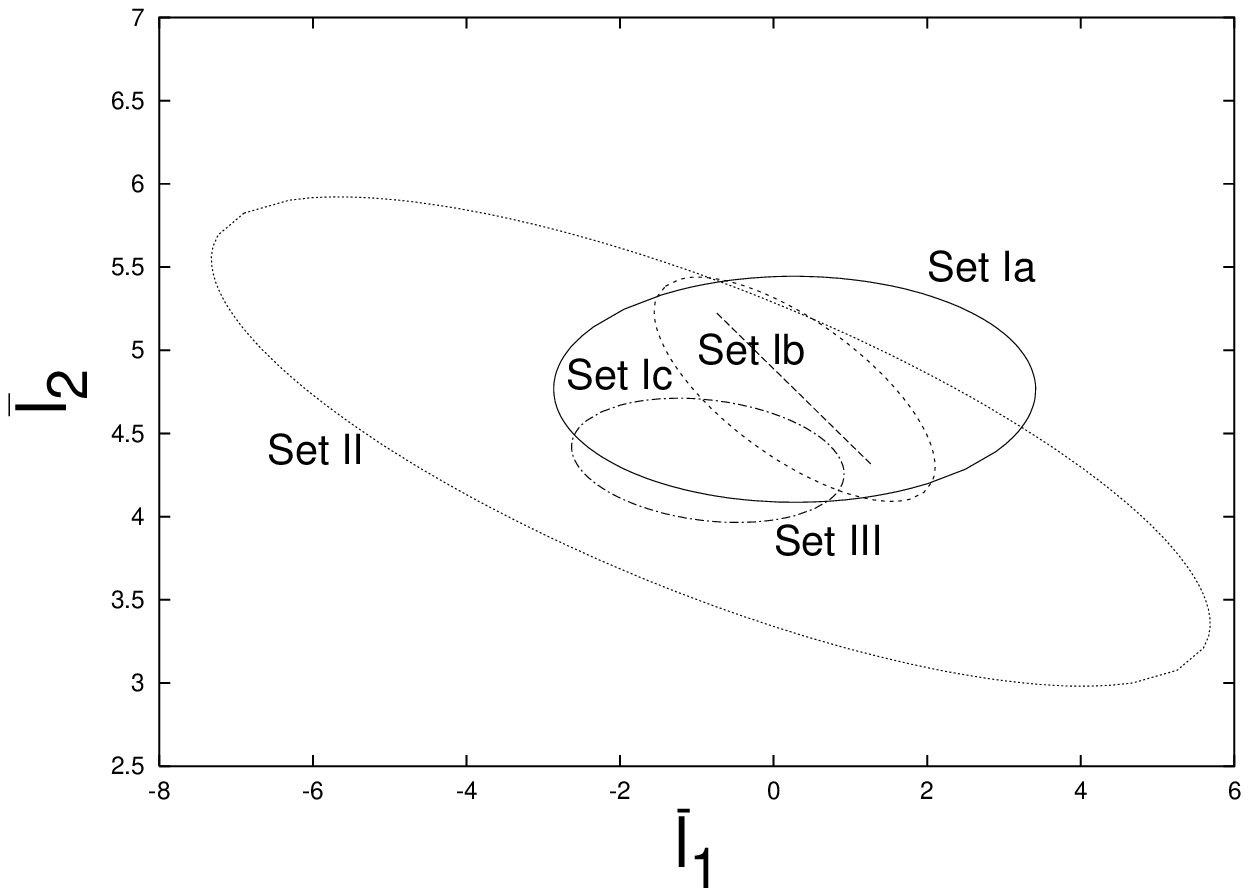,height=6cm,width=8.5cm}}
\caption{ \footnotesize Two dimensional distributions of the low
energy parameters $\bar l_1, \bar l_2 $ generated via Monte Carlo
simulation with $N=10^4$ samples. The contours and the line made of
squares for the Set {\bf Ib} represent the $68\%$ confidence limit, i.e.,
inside them the $68\%$ of the samples are enclosed . Upper left
panel, parameter Sets {\bf Ia} and {\bf Ib} corresponding to a two
loop $K_{l4}$ form factor analysis~\cite{abt99}, with totally
de--correlated and totally anti--correlated $\bar l_1$ and $\bar l_2$
parameters respectively. Upper right panel, parameter Set {\bf Ic}
corresponding to a two loop $K_{l4}$ form factor
analysis~\cite{abt99}, with linear correlation $r=-0.69 $. Middle left
panel, parameter Set {\bf II} corresponding to $D-$waves in $\pi\pi$
scattering ($r=-0.75 $). Middle right panel, parameter Set {\bf III}
corresponding to Roy sum rules in $\pi\pi$ scattering ($r=-0.22
$). Finally, in the bottom panel  the $68\%$
confidence limit contours are superposed.}
\label{fig:eles}
\end{figure}
The determination of the one loop parameters, their corresponding
uncertainties as well as their mutual correlation can be best seen by
plotting the resulting two-dimensional Monte Carlo generated
distributions for all considered sets. This is done in
Fig.~\ref{fig:eles}. For all sets we consider a sufficiently large
sample, $N=10^4$, and have checked numerical stability by doubling the
sample size. We also plot the $68\%$ confidence limit contours, i.e.,
inside these boundaries $68\% $ of distribution points are
enclosed. For a better comparison we have also superposed (bottom
panel) the $68\%$
confidence limit contours to check for statistical compatibility. As
could already be anticipated, the main difference between parameter
Sets {\bf Ia}, {\bf Ic} and {\bf Ib} (corresponding to $K_{l4}$ 
decays) is that the errors decrease as the anti--correlation increases.
The parameter Set {\bf II} provides the largest uncertainties, and
large overlap with the remaining sets is encountered. At the same
time, it is clear that there is a trend to discrepancy between
parameter Sets {\bf Ic} and {\bf III}, since very little overlap is
observed between both distributions. We should also mention that the
level of uncertainty regarding both the $K_{l4}$ and the Roy sum rules
determinations is comparable, although the latter method seems to be
slightly more accurate.

In Ref.~\cite{bc97} the calculation of the two loop contributions to
the threshold parameters was undertaken and explicit expressions for
scattering lengths and ranges were written, in terms of the low energy
constants, $b_{1,2,3,4,5,6} $.  In this note we complete their
numerical calculation by providing their numbers with the inherited
error-bars and also exploring different parameter sets. This is shown
in Table~\ref{tab:bes} (for the  $b_i$parameters) and in
Table~\ref{tab:aes} (for the threshold parameters, $a_{IJ},b_{IJ}$).

For all sets, we have generated statistical distributions both in the
one loop and the two loop corrections to the scattering lengths and
effective ranges by propagating the statistical distributions in the
low energy parameters $\bar l_{1,2,3,4} $, the resonance parameters
$r_{1,2,3,4,5,6} $ and the scale $\mu$ disccused above. The
uncertainty in the scale and in the resonance contributions, affect to
the error of the two loop corrections only. Errors in the two
loop contributions are correlated with those in the one loop
contributions.  Therefore, the error of the sum cannot be obtained by
simply adding the errors in quadrature of one and two loop
contributions.  In any case, for the two loop calculation to be
numerically meaningful these one loop uncertainties have to be
significantly smaller than the corrections due to the two loop
calculation. At the same time the two loop correction has to be
significantly smaller than the one loop correction, for a convergent
expansion.

As we see in Table~\ref{tab:bes}, all predictions for the $b_i$
parameters are compatible within errors. Moreover, as one might have
inferred from the values and distributions of the parameters $\bar l_1
$ and $\bar l_2$ the increasing accuracy of the determinations
correspond to Sets {\bf II}, {\bf Ia}, {\bf Ic}, {\bf Ib} and {\bf
III}, in that order\footnote{These two latter parameter sets have
similar accuracy.}. Besides, we would like to
point out that for the Sets {\bf Ia,Ib} and {\bf Ic} the statistical
fluctuations in the one loop parameters induce errors on the $b_i$
parameters as important, when no more as in the de--correlated case of
Set {\bf Ia}, as the scale uncertainty quoted in Ref.~\cite{bc97}.
For the parameter Set {\bf II} the central values of the $b$
parameters are slightly different than those in Ref.~\cite{bc97}
because the central $\bar l_1 $ and $\bar l_2 $ values are also
different. The errors for the parameter Set {\bf II} are much larger
than those found for parameter Sets {\bf I} and {\bf III}. The size of
the errors for Set {\bf II} is comparable with the ones for $ b_3 ,
b_4 , b_5 , b_6 $ found in Ref.~\cite{mksf95}\footnote{ Regarding the
parameter Set {\bf III}, one should say that if one would calculate
the $b$'s from Eq.~(48) of Ref.~\cite{mksf95}, with the numerical input
of Eq.(2) of  Ref.~\cite{gir97}, one would have obtained 
\begin{eqnarray} 
10 \, b_1 = -0.68 \pm 0.09 \quad , \, 
10 \, b_2 = +0.64  \pm 0.09 \quad , \, 
10^2 \, b_3 = -0.35 \pm 0.24 \nonumber \\ 
10^2 \, b_4 = + 0.47 \pm 0.03 \quad , \, 
10^3 \, b_5 =  0.13 \pm 0.06 \quad , 
10^3 \, b_6 = 0.10 \pm 0.01 \nonumber \nonumber 
\end{eqnarray} 
$b_{3,4,5,6}$ are in agreement with Eq.(49) of Ref.~\cite{mksf95}. We
see that this is {\it not the same} as evaluating the $b$'s in the
spirit of resonance saturation (Eqs. D.1,D.2 and D.3 of second entry
in Ref.~\cite{bc97}) from previously computed values of ${\bar l}_1$
and ${\bar l}_2$, through their relation to $\lambda_1$ and $\lambda_2$
parameters, as suggested in Ref~\cite{gir97} and adopted in this
work. In any case, the values given above and those corresponding to
Set {\bf III} in Table~\protect\ref{tab:bes} are compatible within
statistical fluctuations.}.
\begin{table}[t]
\begin{center}
\begin{tabular}{c|c|c|c|c|c}  
 & Set {\bf Ia} & Set {\bf Ib} &  Set {\bf Ic} &Set {\bf II} & Set {\bf III} 
\\\hline\tstrut
$10 \cdot b_1 $   &$ -0.74\pm 0.21$  &$ -0.74\pm 0.05 $ 
& $-0.74 \pm 0.16  $  &$ -0.83 \pm 0.48 $ &$ -0.83 \pm 0.15 $ \\
$10 \cdot b_2   $ &$ +0.71 \pm 0.17$ &$ +0.71 \pm 0.11 $ 
& $ +0.71 \pm 0.12  $  & $   +0.78 \pm 0.41 $&$ +0.78 \pm 0.10$  \\
$10^2 \cdot b_3 $ &$ -0.15~\er{~0.30}{~0.25}$ &$ -0.15~\er{~0.19}{~0.13} $
& $ -0.15~\er{~0.21}{~0.15}$ & $ -0.27 \pm 0.65 $ &$
-0.27~\er{~0.19}{~0.15}  $  \\
$10^2 \cdot b_4 $ &$  +0.53 \pm 0.07$ &$ +0.53 \pm 0.06 $
&$ +0.53~\er{~0.06}{~0.07}  $ &$  +0.48 \pm 0.13 $ &$ +0.47  \pm 0.03 $ \\
$10^3 \cdot b_5 $ &$  +0.22~\er{~0.19}{~0.24} $ &$ +0.22~\er{~0.09}{~0.14}$ 
& $+0.22~\er{0.11}{~0.16} $ &$  +0.09~\er{~0.26}{~0.35}~$  &$ +0.07~\er{~0.10}{~0.13}$ \\
$10^3 \cdot b_6 $ &$ +0.10~\er{~0.03}{~0.04}$ &$ +0.10~\er{~0.03}{~0.04} $
& $+0.10~\er{0.03}{~0.04}$&$  +0.09~\er{~0.02}{~0.04}$ & $ +0.08~\er{~0.03}{~0.04}$ \\
\end{tabular}
\end{center}
\caption{\footnotesize Low energy scale independent two loop parameters 
and their uncertainties due to the error bars in the
$\bar l's$ one loop parameters and the  uncertainties in both the scale 
and the resonant part of the two loop contribution. Set {\bf Ia} 
corresponds to $K_{l4}$ two loop calculation: 
$ \bar l_1 = 0.3 \pm 2.1 \, ,
  \bar l_2 =  4.77 \pm 0.45 \, ,
  \bar l_3 =  2.9 \pm 2.4 \, ,
  \bar l_4 =  4.4 \pm 0.3  $, 
with $\bar l_1$ and $\bar l_2$ uncorrelated.
Set {\bf Ib} 
corresponds to $K_{l4}$ two loop calculation: 
$ \bar l_1 = 0.3 \pm 1.0 \, ,
  \bar l_2 =  4.77 \pm 0.46 \, ,
  \bar l_3 =  2.9 \pm 2.4 \, ,
  \bar l_4 =  4.4 \pm 0.3  $,
with $\bar l_1$ and $\bar l_2$ totally anti--correlated. 
Set {\bf Ic} 
corresponds to $K_{l4}$ two loop calculation: 
$ \bar l_1 = 0.3 \pm 1.2 \, ,
  \bar l_2 =  4.77 \pm 0.44 \, ,
  \bar l_3 =  2.9 \pm 2.4 \, ,
  \bar l_4 =  4.4 \pm 0.3  $, 
with a correlation coefficient $r( \bar
l_1 , \bar l_2) =  - 0.69 $. 
Set {\bf II} corresponds to a two loop $D-$wave $\pi\pi$ 
scattering lengths calculation (see main text): $  
  \bar l_1 = -0.8 \pm 4.8 \, ,
  \bar l_2 =  4.5 \pm 1.1 \, ,
  \bar l_3 =  2.9 \pm 2.4 \, ,
  \bar l_4 =  4.4 \pm 0.3  $, 
with a correlation coefficient $r( \bar l_1 , \bar l_2) =  - 0.75 $. 
Set {\bf III} corresponds to a Roy equation sum rule analysis of $\pi\pi$ 
scattering at two loops: $ 
  \bar l_1 = -0.9 \pm 1.2 \, ,
  \bar l_2 =  4.34 \pm 0.25 \, ,
  \bar l_3 =  2.9 \pm 2.4 \, ,
  \bar l_4 =  4.4 \pm 0.3  $, with a correlation coefficient $r( \bar
l_1 , \bar l_2) =  - 0.22 $. 
The scale is $\mu = 750 \pm 250 $ MeV  under the resonance saturation
hypothesis, for which a $100\% $ error is assumed. }
\label{tab:bes}
\end{table}

\begin{table}[ht]
\vspace{-1.3cm}
\begin{center}
\begin{tabular}{cc|ccc|c|c} $ a_{IJ} ;  b_{IJ} $ 
& Set & ${\rm (tree)}$ &  ${\rm+  (1 loop)}$&  ${\rm+ (2loop)}$
&  ${\rm total }$&  ${\rm experiment }$
\\\hline\tstrut
$a_{00} m $ 
&{\bf Ia} & $ 0.156  $ & $ 0.043 \pm 0.005$ & $ 0.015 \pm 0.004 $
               & $ 0.214 \pm 0.008$ & $ 0.26 \pm 0.05 $ \\
&{\bf Ib} & $  ``    $ & $\quad ``\quad  \pm 0.002$ & $\quad ``\quad \pm 0.003$
               & $ \quad ``\quad \pm 0.004$ & $  ``   $ \\
&{\bf Ic} &$    ``  $ & $ \quad ``\quad \pm 0.003$ & $ \quad ``\quad \pm 0.003$
               & $ \quad ``\quad \pm 0.005$ & $  ``   $ \\
&{\bf II} & $   ``  $ & $ 0.040 \pm 0.007$ & $ 0.013 \pm 0.004 $
               & $ 0.209 \pm 0.010$ & $  `` $ \\
&{\bf III} &$   ``  $ & $ 0.039 \pm 0.003$ & $ 0.012 \pm 0.002$
               & $ 0.208 \pm 0.005$ & $  ``  $ \\
\hline\tstrut
$b_{00} m^3 $ 
&{\bf Ia} & $ 0.179  $ & $ 0.070 \pm 0.015$ & $ 0.025 \pm 0.011 $
               & $ 0.273 \pm 0.024$ & $ 0.25 \pm 0.03 $ \\
&{\bf Ib} & $   ``     $ & $ \quad ``\quad \pm 0.003$ & 
          $ \quad ``\phantom{````}\phantom{``}\quad \er{~0.007}{~0.008} $
       & $ \quad ``\quad \er{~0.008}{~0.009}    $ & $      ``       $\\
&{\bf Ic} & $   ``     $ & $ \quad ``\quad \pm 0.006$ & 
          $ \quad ``\phantom{``}\quad \pm 0.008 $
       & $ \quad ``\quad \er{~0.011}{~0.012}    $ & $     ``       $\\
&{\bf II} & $    ``    $ & $ 0.059 \pm 0.024$ & $ 0.019 \pm 0.013 $
               & $ 0.257~\er{~0.032}{~0.036} $ & $         ``    $ \\
&{\bf III} & $   ``  $ & $ 0.058 \pm 0.008$ & $ 0.018 \pm 0.006 $
               & $ 0.254 \pm 0.010 $ & $      ``     $ \\
\hline\tstrut
$10 \cdot a_{11} m^3$ &{\bf Ia} & $ 0.297  $ & $ 0.055 \pm 0.012$ & 
             $ 0.021~\er{~0.003}{~0.004}$
               & $ 0.373~\er{~0.011}{~0.012}$ & $ 0.38 \pm 0.02$ \\
         
         &{\bf Ib} & $    ``    $ & $\quad ``\quad \pm 0.009$ & 
             $ \quad ``\quad~\er{~0.002}{~0.003}$
               & $ \quad ``\quad \pm 0.011$ & $      ``        $ \\
         
         &{\bf Ic} & $    ``    $ & $\quad ``\quad \pm 0.009$ & 
             $ \quad ``\quad~\er{~0.002}{~0.003}$
               & $ \quad ``\quad \pm 0.011$ & $      ``        $ \\
         
         &{\bf II} & $   ``     $ & $ 0.059 \pm 0.033$ & 
              $ 0.018~\er{~0.003}{~0.004}$
               & $ 0.375~\er{~0.035}{~0.031}$ & $      ``        $ \\

         &{\bf III} & $   ``     $ & $ 0.059 \pm 0.007$ & $ 0.018 \pm 0.002$
               & $ 0.374 \pm 0.007$ & $      ``       $  \\

\hline\tstrut
$10 \cdot b_{11} m^5$ &{\bf Ia}& $  0    $ & $ 0.030 \pm 0.012$ & $
         0.025~\er{~0.006}{~0.008}$
               & $ 0.055 ~\er{~0.008}{~0.011}$ & $  -   $ \\
         
         &{\bf Ib}& $  ``     $ & $ \quad ``\phantom{``}\quad~\er{~0.008}{~0.009}$ & $
\quad ``\quad ~\er{~0.002}{~0.004}$
               & $ \quad ``\quad~\er{~0.009}{~0.011}$ & $  -   $ \\
         &{\bf Ic}& $  ``     $ & $ \quad ``\quad\pm 0.009$ & $
\quad ``\quad~\er{~0.003}{~0.005}$
               & $ \quad ``\quad \er{~0.009}{~0.011}$ & $  -   $ \\
         
         & {\bf II}& $  ``     $ & $ 0.034 \pm 0.033$ & 
$ 0.020 ~\er{~0.005}{~0.009}$
               & $ 0.054 \pm 0.029$ & $  -   $ \\
         
         & {\bf III}& $  ``     $ & $ 0.034 \pm 0.007$ & 
$ 0.019 ~\er{~0.003}{~0.005}$
               & $ 0.053~\er{~0.005}{~0.006} $ & $  -   $ \\
\hline\tstrut
$10 \cdot a_{20} m $ 
&{\bf Ia} & $ -0.446  $ & $ 0.021 \pm 0.021 $ & $ 0.002 \pm 0.004$
               & $-0.423 \pm 0.024 $ & $-0.28 \pm 0.12$ \\

&{\bf Ib} & $    `` $ & $\quad ``\quad \pm 0.008 $&$ \quad ``\quad\pm 0.004$
               & $\quad ``\quad \pm 0.010 $ & $     ``         $ \\
&{\bf Ic} & $    ``     $ & $ \quad ``\quad \pm 0.011 $ & 
$ \quad ``\quad \pm 0.004 $
               & $\quad ``\quad \pm 0.013 $ & $     ``         $ \\
         
         &{\bf II}& $     ``    $ & $ 0.007 \pm 0.028 $ & 
$ 0.000 \pm 0.004$
               & $-0.439~\er{~0.028}{~0.030} $ & $       ``       $ \\
         
         &{\bf III}& $     ``    $ & $ 0.004 \pm 0.013 $ & 
$ -0.001~\er{~0.003}{~0.004}$
               & $-0.442 \pm 0.015 $ & $       ``       $ \\
\hline\tstrut
$10 \cdot b_{20} m^3 $ &{\bf Ia}& $-0.892  $ & $ 0.129 \pm 0.043$ & 
$ 0.003~\er{~0.013}{~0.011} $
               & $ -0.760 \pm 0.043$ & $ -0.82 \pm 0.08 $ \\
           
         &{\bf Ib}& $   ``     $ & $ \quad ``\quad \pm 0.010$ & 
$ \quad ``\quad \er{~0.012}{~0.010}  $   
            & $ \quad ``\quad \er{~0.016}{~0.014}$ & $      ``          $ \\
         &{\bf Ic}& $   ``     $ & $ \quad ``\phantom{``}\quad~ \er{~0.018}{~0.019}$ & 
$ \quad ``\quad~~ \er{~0.013}{~0.011}  $   
            & $ \quad ``\quad~~~ \er{~0.023}{~0.021}$ & $      ``          $ \\
         
         &{\bf II}& $   ``     $ & $ 0.095 \pm 0.045$ & 
         $ 0.003~\er{~0.016}{~0.013} $
               & $ -0.794\pm 0.039 $& $       ``         $ \\
         
         &{\bf III}& $   ``     $ & $ 0.088 \pm 0.022$ & $ 0.003 \pm 0.011 $
               & $ -0.801\pm 0.021 $& $       ``         $ \\
\hline\tstrut
$10^2 \cdot
 a_{02} m^5 $ &{\bf Ia}   & $  0  $ & $  0.143 \pm 0.031  $ & $ 
0.058~\er{~0.016}{~0.019}$
               & $  0.202 \pm 0.042  $ & $ 0.17 \pm 0.03 $ \\
            
         &{\bf Ib}   & $  ``   $ & $ \quad ``\quad \pm 0.009    $ & $ \quad ``\quad ~\er{~0.016}{~0.019}$
               & $  \quad ``\quad ~ \er{~0.021}{~0.025}  $ & $       ``        $ \\
           
         &{\bf Ic}   & $  ``   $ & $ \quad ``\quad \pm 0.015    $ & $ \quad ``\quad \er{~0.016}{~0.019}$
          & $  \quad ``\quad~ \er{~0.024}{~0.028}  $ & $       ``        $ \\
           
         &{\bf II}  & $  ``  $ & $  0.117 \pm 0.028 $ & $ 0.053 \pm 0.022$
               & $  0.170 \pm 0.030  $ & $       ``        $ \\
           
         &{\bf III}  & $  ``  $ & $  0.111 \pm 0.016 $ & 
           $ 0.051~\er{~0.013}{~0.016}$
               & $  0.163 \pm 0.020  $ & $       ``        $ \\
\hline\tstrut
$10^3 \cdot
 a_{22} m^5 $ &{\bf Ia}& $ 0  $ & $  0.278 \pm 0.241$ & $ 
-0.035~\er{~0.09}{~0.07} $
               & $  0.24 \pm 0.17 $ & $ 0.13 \pm 0.30$ \\
           
         &{\bf Ib}& $ `` $ & $ \quad ``\quad\pm 0.062 $ & $ \quad
               ``\quad~\er{~0.057}{~0.039} $ & $ \quad
               ``\quad~\er{~0.07}{~0.05} $ & $ `` $ \\

         &{\bf Ic}& $  ``  $ & $ \quad ``\quad \pm 0.108 $ & $ \quad
               ``\quad \er{~0.064}{~0.044} $
               & $  \quad ``~\quad \er{~0.10}{~0.08} $ & $       ``       $ \\
           
         &{\bf II}& $  ``  $ & $ 0.119\pm 0.460 $ & $ 0.011~\er{~0.172}{~0.124} $
               & $  0.13\pm 0.30 $ & $       ``       $ \\
           
         &{\bf III}& $  `` $ & $ 0.102 \pm 0.130$   & $ 0.015~\er{~0.072}{~0.061} $
               & $  0.12\pm 0.11 $ & $       ``       $ \\
\end{tabular}
\end{center}
\caption{\footnotesize Threshold  $\pi\pi$ scattering parameters and their
uncertainties in units of $m$ due to the error bars in the
$\bar l's$ one loop parameters and the  uncertainties in both the scale 
and the resonant part of the two loop contribution. Sets {\bf Ia}, 
{\bf Ib}, {\bf Ic}, {\bf II} and {\bf III} are defined in table 1 and
main text. Errors are {\it not} 
added in quadrature due to statistical correlations. 
Experimental values are from Ref.~\protect\cite{Du83}.}
\label{tab:aes}
\end{table}

As we see in Table~\ref{tab:aes}, results for $a_{20}$, $b_{20} $ and
  $a_{22}$ with parameter Set {\bf Ia} are such that the one loop
  errors are larger than the central values of the two loop
  contribution. The inclusion of statistical correlations, Sets {\bf
  Ib} and {\bf Ic}, increases the accuracy of the predictions, though
  it still happens that errors on the one loop contribution make
  irrelevant the two loop one.  The situation worsens dramatically,
  for parameter Set {\bf II} where we see that in most considered
  cases predictive power is lost beyond one loop, with the exception
  of the $S-$wave scattering lengths $a_{00}$ and $a_{02}$.  Set {\bf
  III}, obtained from the Roy-sum rule analysis of Ref,~\cite{gir97},
  turns out to be as predictive as Sets {\bf Ib} and {\bf Ic},
  obtained from the two-loop improved $K_{l4}$ analysis of
  Ref.~\cite{abt99}. In any case is also true that for $a_{20}$,
  $b_{20} $ and $a_{22}$ within Set {\bf III} the one loop
  uncertainties are larger than the two loop contribution.  The
  results of the table are compatible, for all parameter sets with the
  experimental analysis of $\pi\pi$ scattering data~\cite{Du83} but
  produce in general much better errors despite of the problems
  discussed above on the relevance of the two loop contributions. The
  early work on $K_{l4}$ form factors using a dispersive one loop
  calculation for three flavors~\cite{bcg94}, predicted a isoscalar
  $D-$wave three standard deviations above the value extracted from
  the experiment~\cite{gir97}, the upgrade of this calculation to two
  loop accuracy in Ref.~\cite{abt99} have contributed to considerably
  improve such a discrepancy.

\begin{figure}
\centerline{
\epsfig{figure=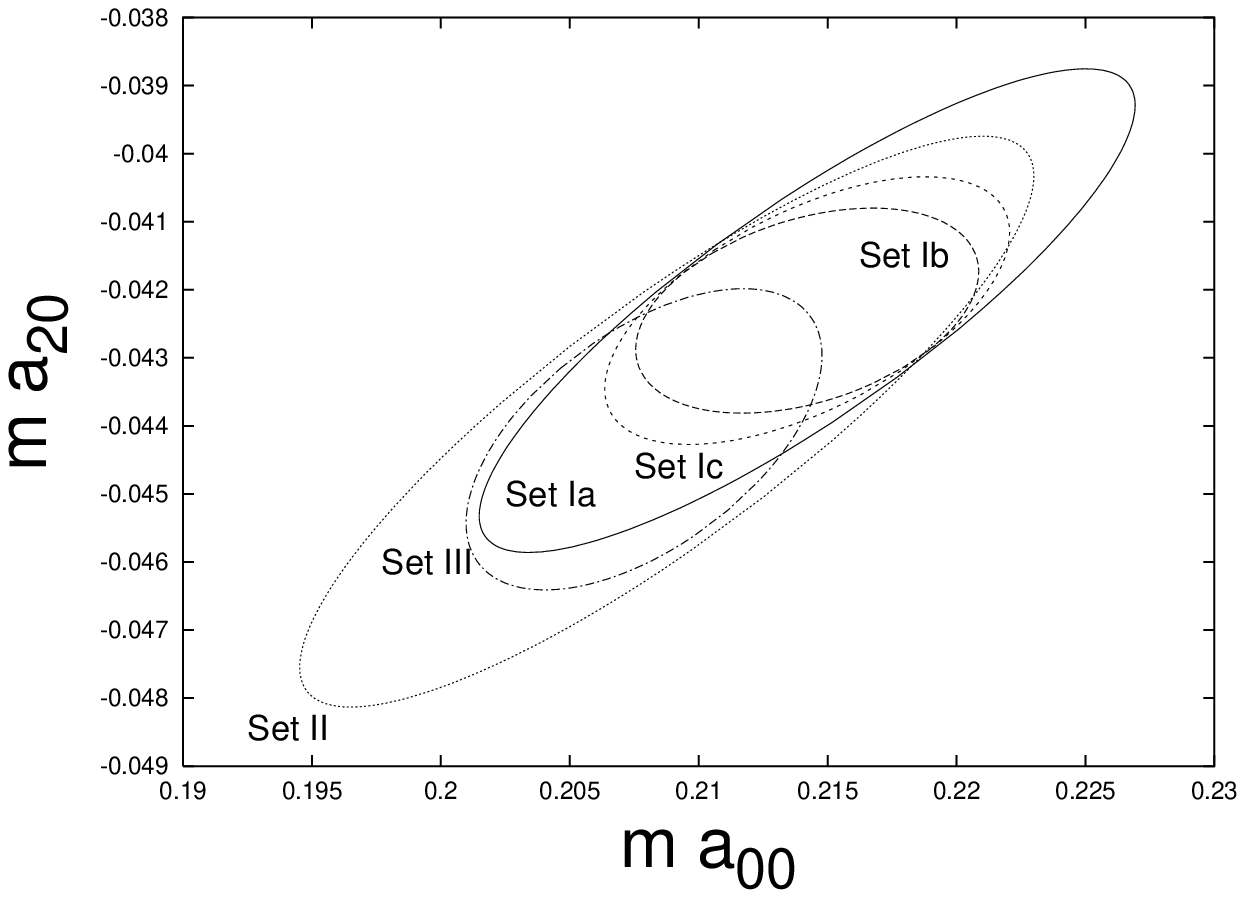,height=7cm,width=8.5cm}
\hspace{.5cm}
\epsfig{figure=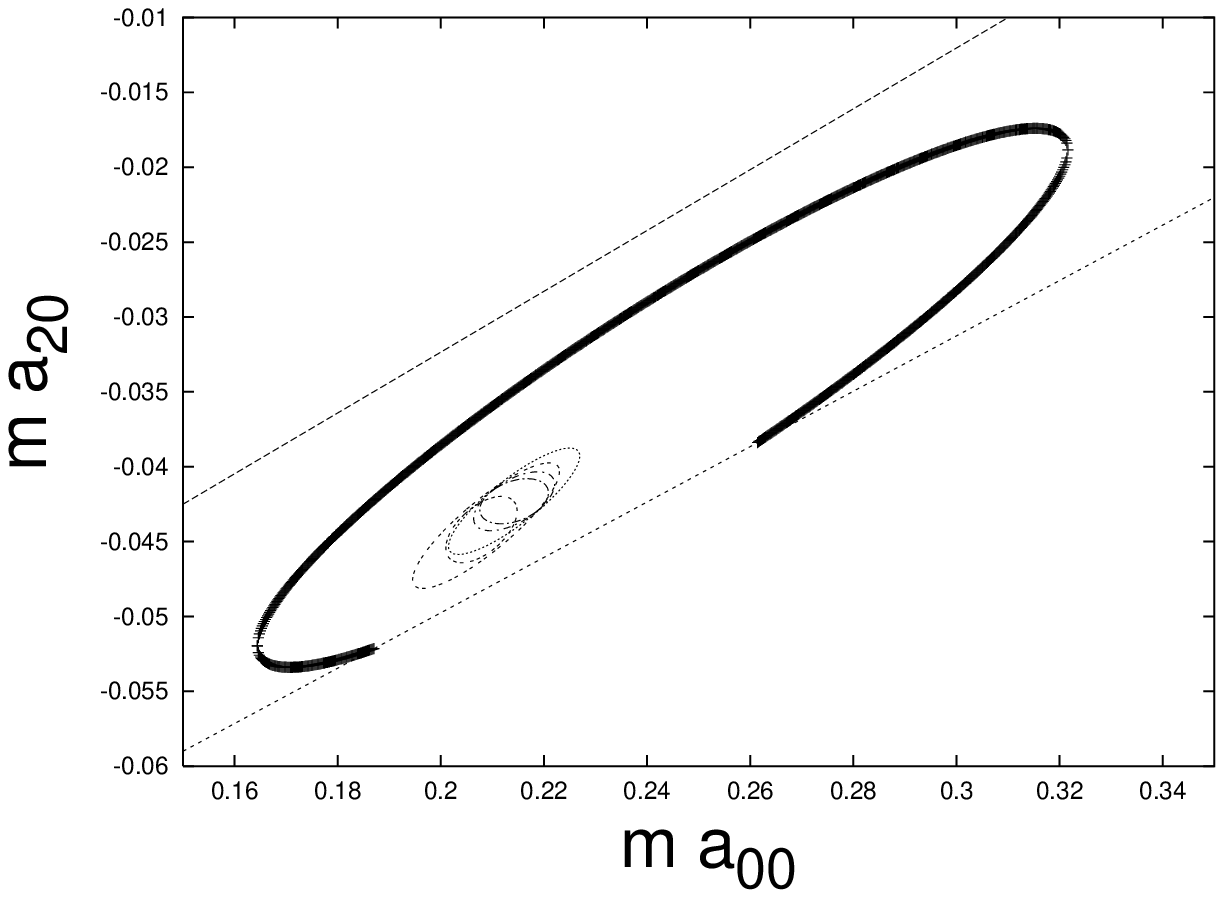,height=7cm,width=8.5cm}}
\caption{ \footnotesize Left panel: contours corresponding to a $68\%$
confidence limits for the two dimensional distributions of the
$S$-wave scattering lengths $a_{00} $ and $a_{20}$. Parameter sets are
defined in the main text. The linear correlation coefficients,
$r(a_{00},a_{20})$, for Sets {\bf Ia, Ib, Ic,II} and {\bf III} are
0.85,0.38,0.58,0.86 and 0.56 respectively. Right panel: The same as
before with the additional inclusion of the universal bands and the
$70\%$ confidence limit contour (crosses) deduced from a numerical
analysis of the Roy equations in Ref.~\protect\cite{col00}.}
\label{fig:aes}
\end{figure}

In Fig.~\ref{fig:aes} we exhibit the Monte Carlo propagated two
dimensional distributions of the $S$-wave scattering lengths, 
for isospin channels
$I=0$ and $I=2$, for all parameter sets and in terms of the $68\%$
confidence limit contours. It is interesting to compare this
distributions with the preliminary results of Ref.~\cite{col00}, based on the
detailed numerical analysis of Roy equations in $\pi\pi$ scattering.  
Such an analysis implies the existence of a universal band, outside
which no solution to Roy equations exists, as a mathematical
consequence of analyticity, crossing, unitarity and isospin
conservation. In addition to this, the physical requirement of 
describing the available experimental data, provides a $70\%$
confidence limit contour~\cite{col00}.  As can be seen in the figure,
with all parameter sets considered in this work, the two-loop ChPT analysis
are not only compatible with the $70\%$
confidence limit contour of Ref.~\cite{col00} but have significant
smaller statistical fluctuations.

In summary, effective field theories like Chiral Perturbation Theory
have  predictive power, but it is not unlimited because of three reasons: 1)
truncation of the expansion, 2) proliferation of undetermined
constants at any order and 3) experimental uncertainties.  Thus,
experimental data prove crucial to determine the, increasing with the
order, unknown constants and their errors, which propagate in a
correlated way to higher orders in the expansion possibly undermining
the ``convergence'' of the expansion. This situation also appears in
fundamental theories like QED or QCD, but it is in fact worse in ChPT
because the number of unknown parameters in these theories does not
increase with the order of the expansion. This why error analysis is
so important. We have exemplified our points in the calculation of the
threshold parameters for $\pi\pi$ scattering up to two loops. The
general picture provided by ChPT is rather satisfactory, in the sense
that the accuracy of the predictions is much bigger than the available
data~\footnote{Right plot of Fig.\ref{fig:aes} constitutes a clear
example.}. Nevertheless we find cases where, within the present experimental
accuracy, the errors in the one loop contribution are larger than the
central values of the two loop contribution. Conclusions regarding the
loss of predictive power, can only be reinforced if systematic
uncertainties induced by the the ${\cal O}(p^6)-$corrections to the
relations between the two-- and three--flavor low energy constants are
included in the results of Ref.~\cite{abt99}. The effect of these
systematic corrections will be twofold: a general increase of the
error fluctuations and a decrease of the correlations. Both effects
will contribute to increase the errors on all derived threshold and
low energy parameters presented in this work and deduced from
$K_{l4}-$decays.

ChPT defines a whole family of effective theories, 
but obviously the most interesting choice exactly corresponds to low 
energy QCD. In QCD with two flavors and neglecting isospin breaking 
there appear only two independent parameters, 
$\Lambda_{\rm QCD}$ and the quark mass $m_Q $. The quantities $f_\pi $, 
$m$, the $\bar l$'s, the $b$'s and higher order parameters must be 
functions of them. 
This dependence introduces correlations among all low energy parameters which,
if taken into account, would influence the present error analysis, and 
presumably might yield to more moderate errors in terms of $\Lambda_{\rm QCD}$ 
and $m_Q$ and their errors. Assuming more statistically independent 
parameters than QCD suggests is, so 
far, another manifestation of the inability to undertake  a quantitative and
microscopic derivation of ChPT as an effective low energy theory of QCD on 
the one hand, but on the other hand reassures ChPT as a convenient tool to 
deal with non-perturbative phenomena in strong interactions. 

\vspace{1cm}

{ \large \bf Appendix: Modeling statistical correlations in $K_{l4}$ decays.} 

\vspace{1cm}

In Ref.~\cite{abt99} is stated that $L_3^r$ is strongly
anti--correlated with $L_1^r$ and $L_2^r$, though central values and
errors for these parameters are given ($L_1^r = 0.52 \pm 0.23$, $L_2^r
= 0.72 \pm 0.24$ and $L_3^r = -2.69 \pm 0.99$, main fit in Table 1 of
Ref.~\cite{abt99}), however the correlation matrix is not provided.
Besides, new parameters $X_1 = L_2^r - 2L_1^r-L_3^r$, $X_2 = L^2_r$
and $X_3 = (L_2^r-2L_1^r)/L_3^r $ are introduced, it is quoted the
value $X_3 =0.12\er{8}{11}$ and it is also specified that $X_3$ is
little correlated with $X_1$ and $X_2$.  We will show below how
statistical modeling and the fragmentary information given in
Ref.~\cite{abt99} can be used to reconstruct some relevant information
on the correlations between $\bar l_1$ and $\bar l_2$. To be most
objective, we also explore two other extreme cases which correspond to
total de--correlation and total anti--correlation. This said, we
explore three different scenarios in this work:

\begin{itemize}

\item A total de--correlated picture ($ r ( L_1^r , L_3^r ) = r ( L_2^r
, L_3^r ) = r ( L_1^r ,L_2^r ) = 0 $, being $r$ the linear correlation
coefficient). It leads to $X_3 = 0.12\er{23}{20}$, $r(X_1,X_3) = -0.14$ and
$ r(X_2,X_3) = -0.09$. 

\item A total anti--correlation scenario ($ r ( L_1^r , L_3^r ) = r (
L_2^r , L_3^r ) = -1 $) which can be implemented by simply assuming a
linear relation
\begin{equation}
 \left (L_1^r - \langle L_1^r \rangle \right)/ \sigma_1 =-
\left (L_3^r - \langle L_3^r \rangle \right) / \sigma_3 = \left (L_2^r
- \langle L_2^r \rangle \right ) / \sigma_2,  \label{eq:corr}
\end{equation}
That is the necessary and sufficient condition for total
anti--correlation. Besides, one also has $ r ( L_1^r , L_2^r ) =
+1$. This simple model leads to $X_3 = 0.12\er{3}{6}$ and $r(X_1,X_3) =
r(X_2,X_3) = 0.06$. Thus, the total anti--correlation scenario, though
simple, provides an acceptable description, and in any case much more
precise than when correlations between $L_1^r$, $L_2^r$ and $L_3^r$
are neglected, of the findings of Ref.~\cite{abt99}.

\item A partial anti--correlation scenario  which can be
implemented  if one assumes  that  $L^r_1$ , $L^r_2$ and $L^r_3$
are gaussian distributed\footnote{This is totally justified, because
in Ref.~\protect\cite{abt99}  these variables  have been determined from
a $\chi^2-$fit.} according to 
\begin{eqnarray}
P(L_1^r,L_2^r,L_3^r) &=& \frac{\left({\rm det}[C]\right)^\frac12
}{(2\pi)^\frac32\sigma_1\sigma_2\sigma_3} \times
e^{-\frac12 \left( L^T\cdot C\cdot L \right) }\label{eq:Cmatrix}
\end{eqnarray}
where $P$ is the join density probability distribution of the three
random variables, $\sigma_i$ is the $L_i^r$ error, $L$ is a column
matrix with elements $(L_i^r-\mu_i)/\sigma_i$, being $\mu_i$ the
central value of the variable $L_i^r$, and finally the inverse of the
symmetric matrix $C$ is given by
\begin{equation}
 C^{-1}= \left (
\begin{array}{ccc}1&r_{12}&r_{13}\\r_{12}&1&r_{23}\\r_{13}&r_{23}&1
\end{array}\right)
\end{equation}
and therefore it is determined by  the linear correlation coefficients $r_{
ij} = r ( L_i^r , L_j^r )$. 
Our aim is to improve the total anti--correlated scenario presented in
the item above but keeping the model still as simple as possible. Thus
we have explored situations where $r_{13} = r_{23}=-r_{12} = r$, for
which one can analytically diagonalize the $C-$matrix in
Eq.~(\ref{eq:Cmatrix}).  We find for $r=-0.85$, still a strong
anti--correlation, $X_3 = 0.12\er{8}{10}$, $r(X_1,X_3) = -0.02 $ and
$r(X_2,X_3) = -0.01$. Thus, this model for the correlations reproduces
pretty much the results given in the $K_{l4}$ analysis of
Ref.~\cite{abt99}.  This can also be appreciated in
Fig.~\ref{fig:L12L3} where we show the resulting two dimensional distributions
of the $SU(3)-$low energy parameters $L_1^r, L_3^r$ and $L_2^r,
L_3^r$. These distributions compare reasonably well with those given
in Fig.2 of the first entry of Ref.~\cite{abt99}. This makes us more
confident on the validity of the simple statistical model used here.

\begin{figure}
\centerline{
\epsfig{figure=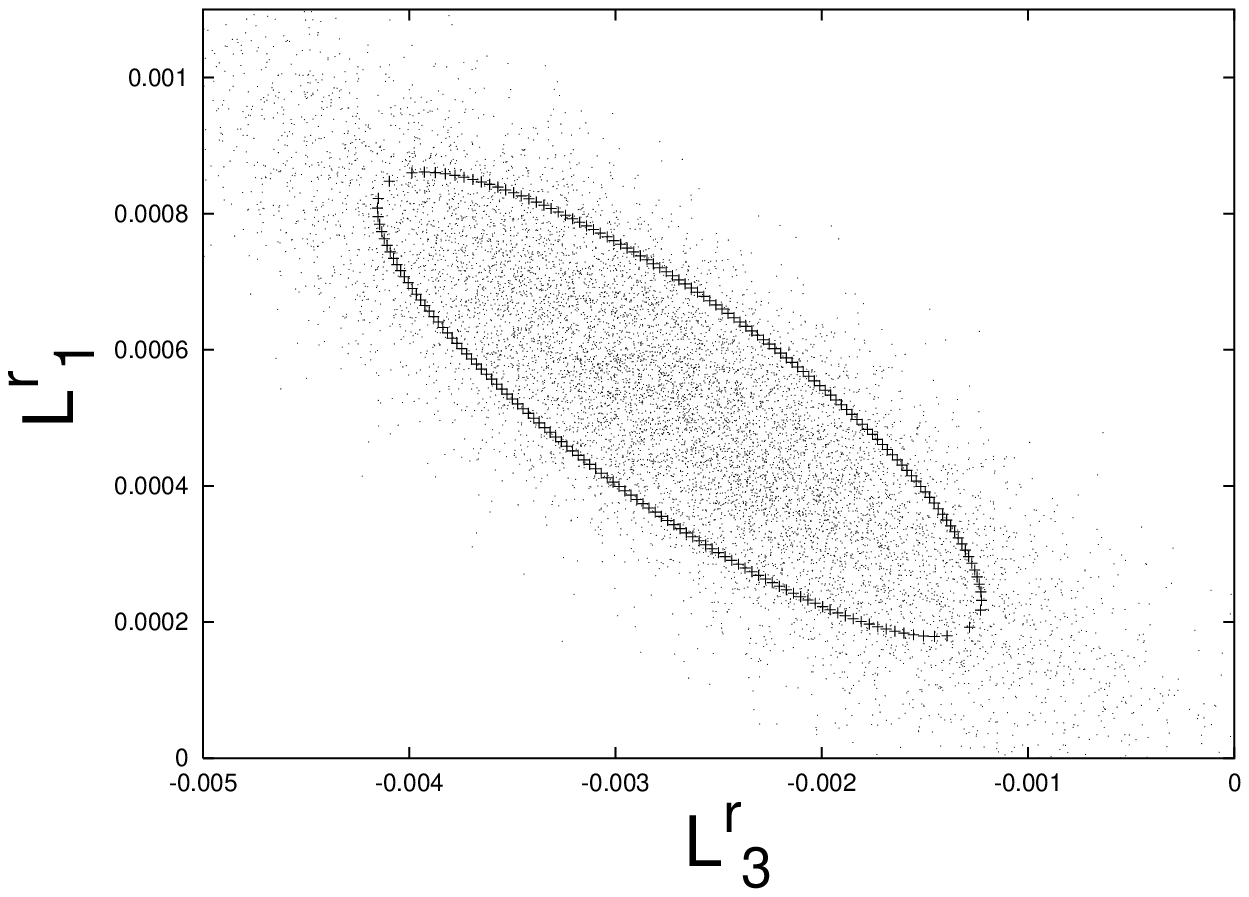,height=7cm,width=8.5cm}
\hspace{.5cm}
\epsfig{figure=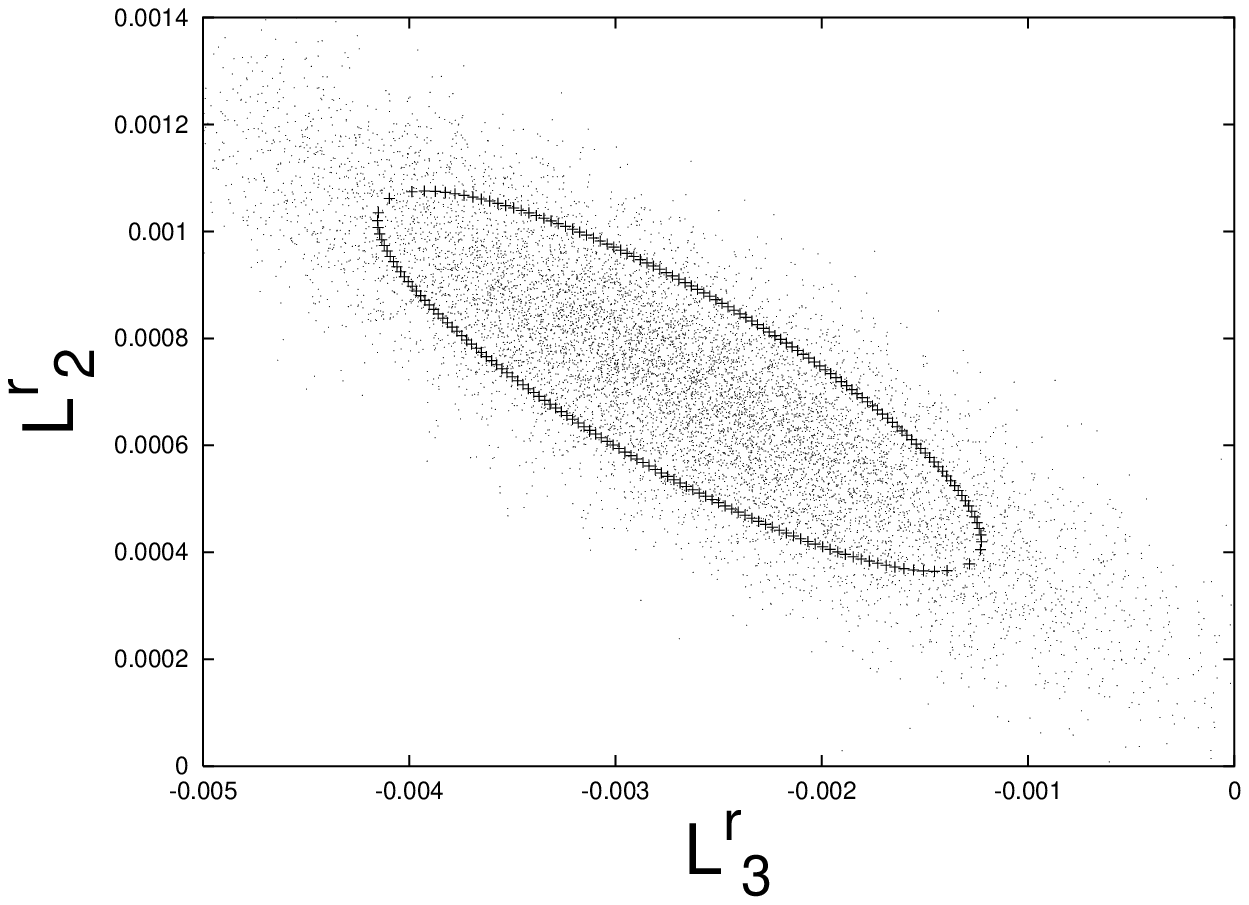,height=7cm,width=8.5cm}}
\caption{ \footnotesize Two dimensional distributions of the
$SU(3)-$low energy parameters $L_1^r, L_3^r$ and $L_2^r,
L_3^r$. generated via Monte Carlo simulation with $N=10^4$
samples. The contours represent the $68\%$ confidence limit, i.e.,
inside them the $68\%$ of the samples are enclosed. }
\label{fig:L12L3}
\end{figure}

\end{itemize}

The correlations between $L_1^r$, $L_2^r$ and $L_3^r$ are transported
through the equations relating $SU(2)$ and $SU(3)$ low energy
constants. For the total anti--correlated scenario leads to a total
anti--correlation for ${\bar l_1}$ and ${\bar l_2}$, it is to say 
$ ( \bar l_1 - \langle \bar l_1 \rangle ) / \sigma_1 =- (\bar l_2 -
\langle \bar l_2 \rangle ) / \sigma_2  $   and therefore $ r ( \bar
l_1 , \bar l_2 ) = -1 $. For the total de--correlated case the
correlation coefficient $ r ( \bar
l_1 , \bar l_2 ) $ is zero, whereas for the partial correlation scenario 
we get $r ( \bar l_1 , \bar l_2 ) = -0.69$.


\begin{thebibliography}{99}
\bibitem{We66} S. Weinberg, Phys. Rev. Lett {\bf 17} (1966) 616,

\bibitem{GL84} J. Gasser and H. Leutwyler, Ann. Phys. (N.Y.) {\bf 158} (1984) 
142. 

\bibitem{mksf95} M. Knecht, B. Moussallam, J. Stern and N.H. Fuchs,
Nucl. Phys. {\bf B457} (1995)513;
{\it ibidem} {\bf B471} (1996)445;


\bibitem{bc97} J. Bijnens, G. Colangelo, G. Ecker, J. Gasser and
M.E. Sainio, Phys. Lett. {\bf B374} (1996) 210; {\it ibidem}
Nucl. Phys. {\bf B508} (1997) 263.

\bibitem{egpr89} G. Ecker, J. Gasser, A. Pich and E. de Rafael,
Nucl. Phys. {\bf B321} (1989) 311.


\bibitem{bcg94} J. Bijnens, G. Colangelo and J. Gasser, Nucl. Phys.
{\bf B 427} (1994) 427.

\bibitem{abt99} G. Amoros, J. Bijnens and P. Talavera, {\it
hep-ph/9912398}; {\it ibidem}  {\it hep-ph/0003258}.

\bibitem{gir97} L. Girlanda, M. Knecht, B. Moussallam an J. Stern,
Phys. Lett. {\bf B 409} (1997) 461.

\bibitem{bct98} J. Bijnens, G. Colangelo and P. Talavera, JHEP 
{\bf 9805} (1998) 014.

\bibitem{gl85} J. Gasser and H. Leutwyler, Nucl. Phys. {\bf B250}
(1985) 465.

\bibitem{Du83} O. Dumbrajs et al., Nucl. Phys. {\bf B 216} (1983)
277. 

\bibitem{col00} G. Colangelo, {\it hep-ph/0001256. }



\end{thebibliography}
\end{document}